\documentclass[10pt,twocolumn]{article} 
\usepackage{acticlestyles}
\usepackage{times}
\usepackage{amssymb}
\usepackage{url,hyperref}
\usepackage[utf8x]{inputenc} 
\newcommand{\ignore}[1]{}
\usepackage{tikz}
\usetikzlibrary {arrows}
\usetikzlibrary{arrows.meta}
\usepackage[export]{adjustbox}
\usepackage{graphicx}
\usetikzlibrary{fit}
\usepackage{textgreek}
\usepackage[hang]{footmisc}
\usepackage{mathrsfs,amsmath}
\usepackage{pifont}
\usepackage{booktabs}
\usepackage{graphicx}
\usepackage{array}
\usepackage{cleveref}

\usepackage{subcaption}
\usepackage[textwidth=11mm]{todonotes}
\begin{document}

\title{Time-encoded mid-infrared Fourier-domain optical coherence tomography}

\author{\textbf{Ivan Zorin\textsuperscript{1,*}, Paul Gattinger\textsuperscript{1}, Andrii Prylepa\textsuperscript{1}, Bettina Heise\textsuperscript{1}} \\
\\
Research Center for Non-Destructive Testing, Science Park 2,\\
Altenberger Str.69, 4040 Linz, Austria\\
\today
\\
*{\normalsize Corresponding author: \href{mailto:ivan.zorin@recendt.at}{ivan.zorin@recendt.at}}\\
\hline
\\
\normalsize
Accepted manuscript; Ivan Zorin, Paul Gattinger, Andrii Prylepa, and Bettina Heise, \\\normalsize“Time-encoded mid-infrared Fourier-domain optical coherence tomography,” Opt. Lett. 46, (2021). \\\normalsize \href{https://doi.org/10.1364/OL.434855}{https://doi.org/10.1364/OL.434855} \\\normalsize © 2021 Optical Society of America
}

\maketitle
\thispagestyle{empty}

\begin{abstract}

We report on a technically simple approach to achieve high-resolution and high-sensitivity Fourier-domain OCT imaging in the mid-infrared range. The proposed OCT system employs an InF\textsubscript{3} supercontinuum source. A specially designed dispersive scanning spectrometer based on a single InAsSb point detector is employed for detection. The spectrometer enables structural OCT imaging in the spectral range from 3140~nm to 4190~nm with a characteristic sensitivity of over 80~dB and an axial resolution below 8~μm. The capabilities of the system are demonstrated for imaging of porous ceramic samples and transition-stage green parts fabricated using an emerging method of lithography-based ceramic manufacturing. Additionally, we demonstrate the performance and flexibility of the system by OCT imaging using an inexpensive low-power (average power of 16~mW above 3~μm wavelength) mid-IR supercontinuum source.

\end{abstract}

\noindent\makebox[\linewidth]{\rule{\columnwidth}{0.4pt}}
\vspace{2pt}

Optical coherence tomography (OCT) is a well-established non-invasive and non-contact method for three-dimensional structural imaging of complex specimens. OCT is of major importance and is primarily used in biomedical scenarios and ophthalmology~\cite{fujimoto_development_2016}. In the last few years, prospects and effectiveness of OCT beyond biomedicine~\cite{Stifter2007} have markedly increased in terms of applied research, as the technique has been pushed to longer wavelengths~\textendash~first, to the border of the near-infrared (near-IR) range~\cite{Cheung:14,Cheung:15}, then further into the mid-IR region. 
The reports on mid-IR OCT (either implemented in direct sensing mode based on thermal detection~\cite{Zorin:18,Zorin_OE:20} or employing nonlinear conversion techniques~\cite{Israelsen:19,Vanselow:20}) have demonstrated advanced capabilities of long wavelengths for probing scattering media. Since the magnitude of scattering decreases with increasing wavelength, it has become possible to examine samples previously ineligible for testing. In this way, these reports have demonstrated the benefits of shifting to the mid-IR spectral region and have identified application opportunities, but also challenges.
%~\cite{Cheung:14,Cheung:15}

As a result, mid-IR OCT has become an emerging sub-field that has already gained significant attention in the field of non-destructive testing and defectoscopy of strongly scattering industrial samples (with no water content, due to strong water absorption in this range). The main obstacle that still severely limits the effectiveness and widespread adoption of OCT operating in this spectral band is essentially technical. It is caused by lagging mid-IR detection technologies or their technical complexity.
It is particularly interesting that the presented mid-IR OCT systems demonstrated enhanced probing capabilities despite severely limited sensitivities (from approximately 60~dB to 80~dB) compared to well-developed near-IR OCT counterparts. Thus, the primary reason was to establish mid-IR OCT in the Fourier-domain (FD-OCT) configuration that has a fundamental sensitivity advantage over the time-domain OCT (TD-OCT)~\cite{Leitgeb:03,Choma:03}. In the former case, all depth components are captured simultaneously and do not contribute to depth-dependent noise as inherent in TD-OCT. %~\cite{Leitgeb:03,Choma:03}

%However, since mid-IR swept-sources are not yet technically mature for OCT, 
The FD-OCT configuration poses a number of realization complications, when transferring to the mid-IR range (mid-IR swept-sources are not yet technically mature for OCT). For example, unlike TD-OCT, which uses a single-point detector, the FD-OCT requires a broadband mid-IR spectrometer with well-separated spectral channels, based on a linear array in the standard configuration. State-of-the-art mid-IR focal plane arrays exhibit compromised sensitivity (inherent for low bandgap semiconductors as well as for thermal detectors). These arrays are expensive (in particular photonic systems) and have small numbers of pixels (thermal systems). For this reason, the reported mid-IR OCT realizations either exploited nonlinear conversion techniques to perform detection using well-developed near-IR technologies, or employed direct sensing using a pyroelectric array with proper handling of the resulting limitations. In the first case, the sensitivity of the systems is strongly affected by the relative inefficiency of the conversion techniques. However, broad spectral ranges and high-performance cameras enabled a high axial resolution and a reasonable sensitivity roll-off. In contrast, direct sensing OCT based on thermal arrays offered higher sensitivity, but had lower axial resolution and a strong trade-off between spatial performance and signal roll-off. In  this case, the spectral range could not be broadened, which led to a sharp drop of sensitivity over depth, as the spectral resolution of the spectrometer is severely limited by the number of pixels.

In this letter, we address these issues and report on a new approach of mid-IR time-encoded FD-OCT (teFD-OCT)~\cite{Drexler,Povazay:06} that not only provides high sensitivity and axial resolution, but is also technically simple and cost-effective. We present a detection system well suited for mid-IR OCT in the current state of the detection technology. This solution elegantly combines the advantages of TD- and FD-OCT by using a single-point detector whilst achieving the sensitivity advantage of FD-OCT~\cite{Povazay:06}.

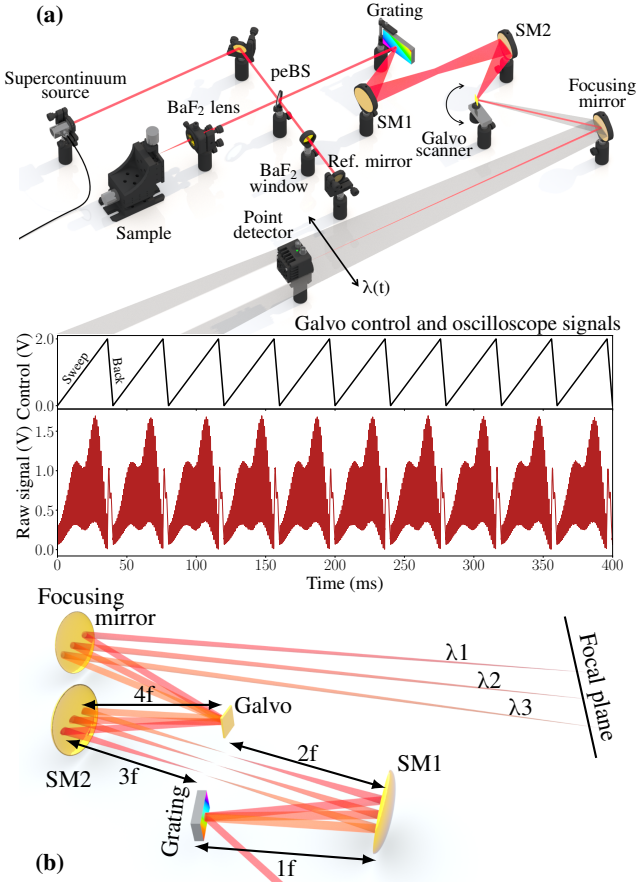
\begin{figure}[!t]
\begin{tikzpicture}
\centering
 \node[anchor=south west,inner sep=0] (image) at (0,0,0) {\includegraphics[width=1\linewidth]{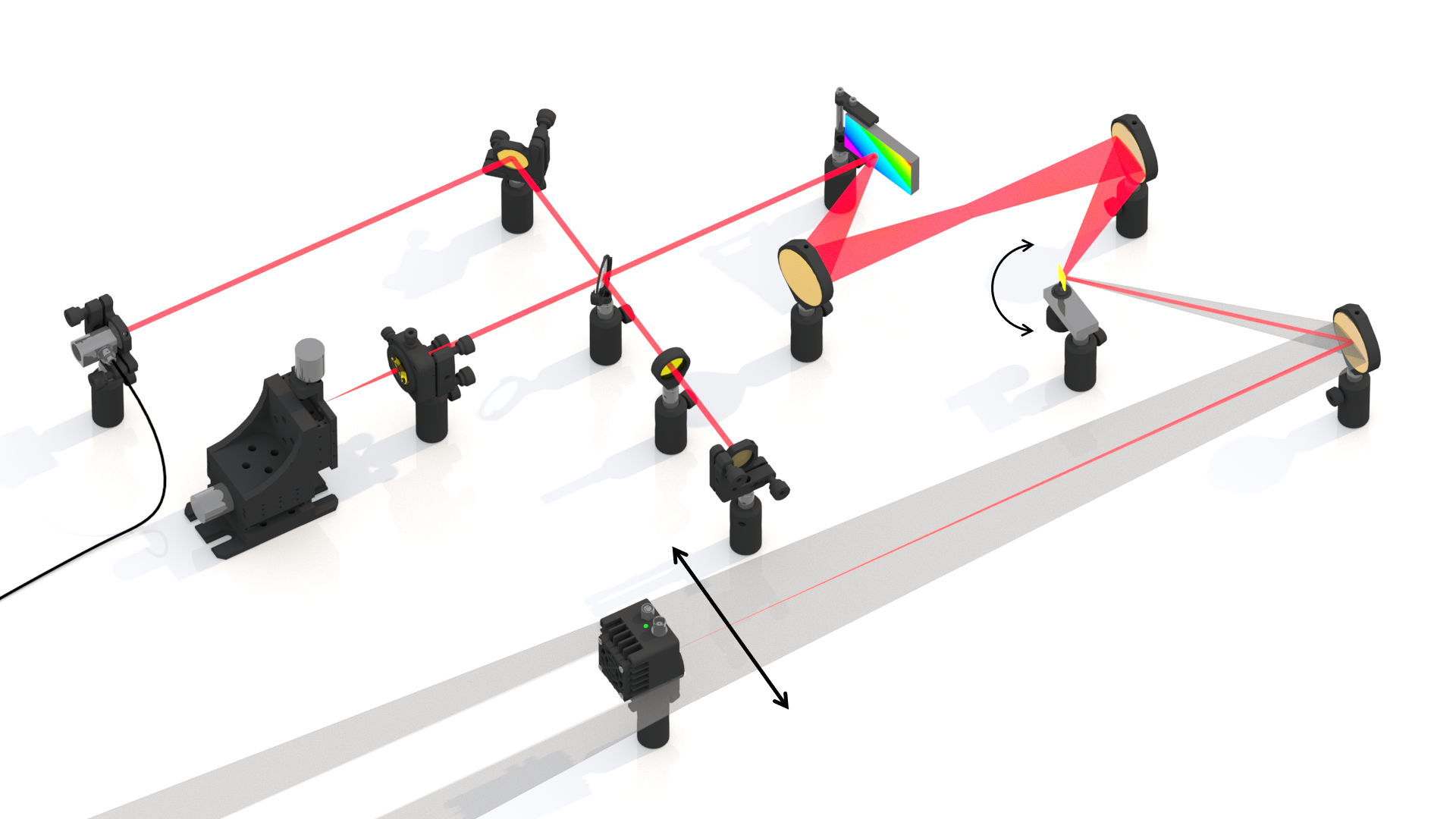}};
 \begin{scope}[x={(image.south east)},y={(image.north west)}]
 %% next four lines will help you to locate the point needed by forming a grid. comment these four lines in the final picture.↓
  \draw (0.075,0.72) node[]{\color{black}\scriptsize Supercontinuum};
  \draw (0.075,0.68) node[]{\color{black}\scriptsize source};
  \draw (0.434,0.74) node[]{\color{black}\scriptsize peBS};
  \draw (0.295,0.64) node[]{\color{black}\scriptsize BaF\textsubscript{2} lens};
  \draw (0.415,0.46) node[]{\color{black}\scriptsize BaF\textsubscript{2}};
  \draw (0.415,0.42) node[]{\color{black}\scriptsize window};
  \draw (0.56,0.49) node[]{\color{black}\scriptsize Ref. mirror};
  \draw (0.6,0.91) node[]{\color{black}\scriptsize Grating};
  \draw (0.6,0.6) node[]{\color{black}\scriptsize SM1};
  \draw (0.82,0.86) node[]{\color{black}\scriptsize SM2};
  \draw (0.678,0.555) node[]{\color{black}\scriptsize Galvo};
  \draw (0.678,0.515) node[]{\color{black}\scriptsize scanner};
  \draw (0.93,0.7) node[]{\color{black}\scriptsize Focusing};
  \draw (0.93,0.66) node[]{\color{black}\scriptsize mirror};
  \draw (0.39,0.33) node[]{\color{black}\scriptsize Point};
  \draw (0.39,0.29) node[]{\color{black}\scriptsize detector};
%     \draw (0.35,0.32) node[]{\color{black}\scriptsize Point detector};
%   \draw (0.35,0.29) node[]{\color{black}\scriptsize with slit};
  \draw (0.57,0.13) node[]{\color{black}\scriptsize \textlambda(t)};
  \node[anchor=south west,inner sep=0] (image) at (0.02,-0.69,0) {\includegraphics[width=0.95\linewidth]{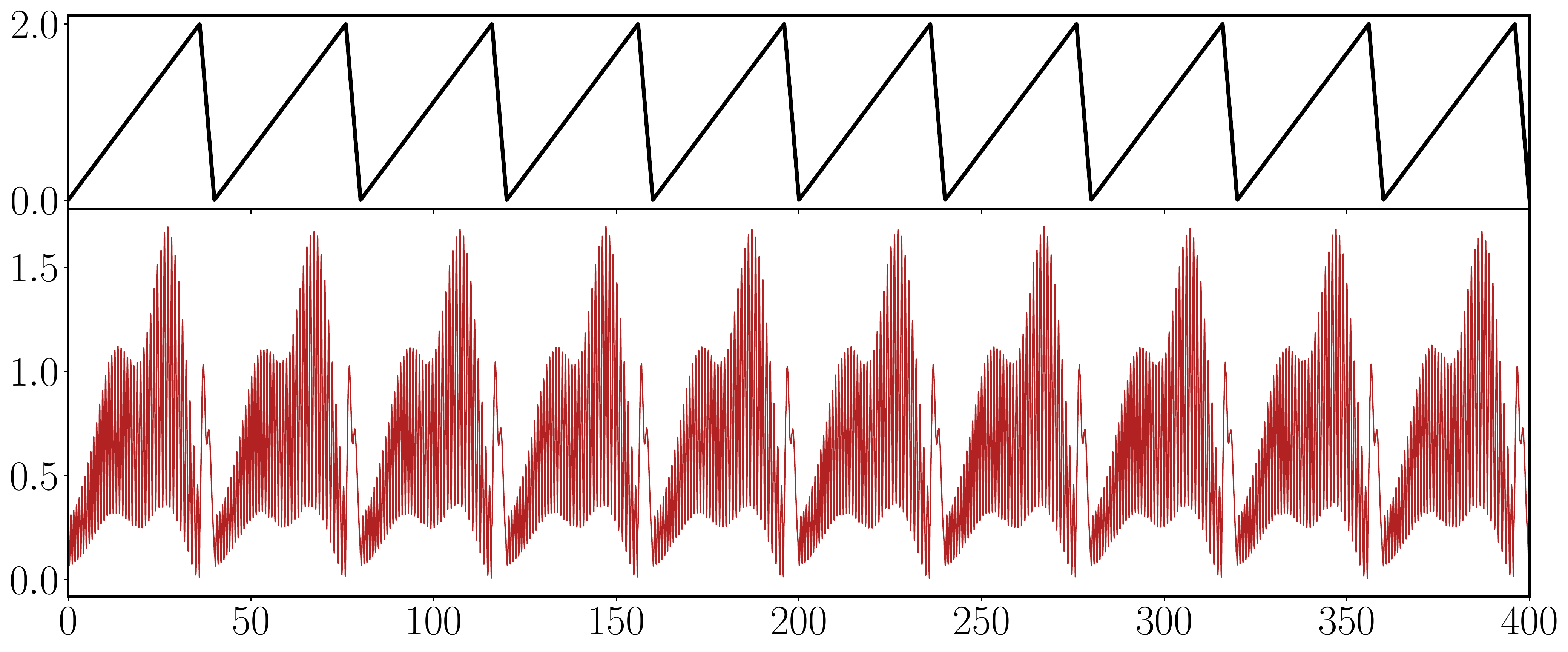}};
  \draw (0.52,-0.71) node[]{\color{black}\scriptsize Time (ms)};
  \draw (0.01,-0.12) node[rotate=90]{\color{black}\scriptsize Control (V)};
  \draw (0.01,-0.42) node[rotate=90]{\color{black}\scriptsize Raw signal (V)};
 \node[anchor=south west,inner sep=0] (image) at (0,-1.55,0) {\includegraphics[width=0.95\linewidth]{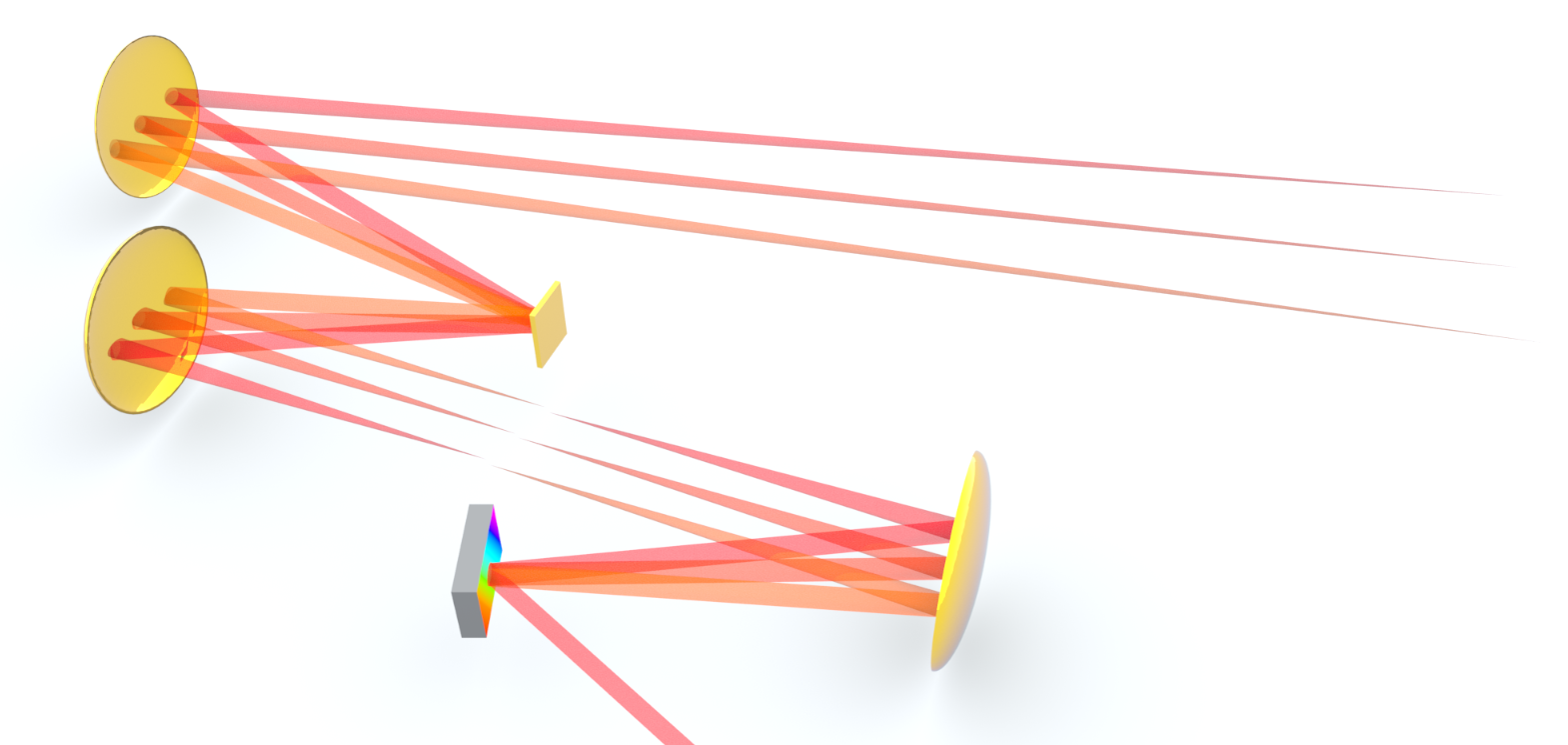}};
  \draw (0.05,0.9) node[]{\color{black}\small \textbf{(a)}};
  \draw (0.05,-1.5) node[]{\color{black}\small \textbf{(b)}};
    \draw [latex-latex,thick, color={rgb,255:red,0; green,0; blue,0}] (0.28,-1.45) to (.57,-1.49); 
    \draw [latex-latex,thick, color={rgb,255:red,0; green,0; blue,0}] (0.585,-1.285) to (.335,-1.157); 
    \draw [latex-latex,thick, color={rgb,255:red,0; green,0; blue,0}] (0.285,-1.27) to (.075,-1.15); 
    \draw [latex-latex,thick, color={rgb,255:red,0; green,0; blue,0}] (0.1,-1.05) to (.325,-1.05); 
     \draw (0.425,-1.51) node[]{\color{black}\small 1f};
     \draw (0.46,-1.185) node[]{\color{black}\small 2f};
     \draw (0.175,-1.25) node[]{\color{black}\small 3f};
     \draw (0.2,-1.018) node[]{\color{black}\small 4f};
     \draw (0.64,-1.215) node[]{\color{black}\small SM1};
     \draw (0.08,-1.26) node[]{\color{black}\small SM2};
     \draw (0.25,-1.38) node[rotate=83]{\color{black}\small Grating};
     \draw (0.1,-0.75) node[]{\color{black}\small Focusing};
     \draw (0.17,-0.8) node[]{\color{black}\small mirror};
     \draw (0.39,-1.05) node[]{\color{black}\small Galvo};
    \draw [-,thick, color={rgb,255:red,0; green,0; blue,0}] (0.87,-0.8) to (.92,-1.2); 
      \draw (0.92,-1) node[rotate=-78]{\color{black}\small Focal plane};
      \draw (0.7,-0.9) node[]{\color{black}\small \textlambda 1};
      \draw (0.75,-0.98) node[]{\color{black}\small \textlambda 2};
      \draw (0.8,-1.055) node[]{\color{black}\small \textlambda 3};
      \draw (0.7,0.025) node[]{\color{black}\footnotesize Galvo control and oscilloscope signals};
     \draw (0.092,-0.098) node[rotate=55]{\color{black}\tiny Sweep};
     \draw (0.16,-0.105) node[rotate=-83]{\color{black}\tiny Back};
     \draw (0.2,0.28) node[]{\color{black}\scriptsize Sample};
 %% upto here↑
 %%
%  \draw[help lines,xstep=.1,ystep=.1] (0,-1.5) grid (1,1);
%  \draw[help lines,xstep=.05,ystep=.05] (0,-1.5) grid (1,1);
%  \foreach \x in {0,1,...,9} { \node [anchor=north] at (\x/10,0) {0.\x}; }
%  \foreach \y in {-15,-14,...,9} { \node [anchor=east] at (0,\y/10) {0.\y};}
 \end{scope}
\end{tikzpicture}
\caption{Basic scheme (simplified) and operational principles of the mid-IR teFD-OCT; (a) optical layout of the OCT system, peBS - pellicle beam splitter; SM1 and SM2 - identical spherical mirrors forming a 4f system to conjugate the grating and scanning planes; raw asymmetric galvo control signals and corresponding real measured spectral interferograms (for a 16~mW mid-IR supercontinuum source) are shown below; (b)~detailed schematic optical diagram of the spectrometer.}
\label{fig:schema}
\end{figure}

\ignore{The developed mid-IR teFD-OCT system was established using standard~\textendash~for an IR laboratory~\textendash~on-shelf components.}
The basic optical layout and operating principles of the mid-IR teFD-OCT imager are shown in Fig.~\ref{fig:schema}. The system is free-space, based on a Michelson interferometer [see Fig.~\ref{fig:schema}(a)] that is formed by a pellicle beamsplitter (nitrocellulose membrane-based, BP145B4, Thorlabs). A 1-inch gold flat mirror is employed in the reference arm. In the sample arm of the interferometer, the supercontinuum beam is focused onto the sample (fixed on a three-axis motorized stage) using a standard BaF\textsubscript{2} singlet lens (6~mm thick, anti-reflection coated, 50~mm focal length, no strong influence of chromatic aberration observed). Due to the significantly wider operational spectral bandwidth accessed by the developed spectrometer, an effect of group velocity dispersion degrading the axial resolution was clearly observed. The lens is a singlet, thus, correction of the sample arm dispersion was done by inserting a BaF\textsubscript{2} flat window (5~mm thick, anti-reflection coated) in the reference arm. The scheme of the interferometer and focusing optics has not changed from the one demonstrated in~\cite{Zorin:20}, hence the characterization of lateral resolution (39~µm at 4~\textmu m wavelength) is not given here.

For the principal design of the key assembly, i.e. the dispersive scanning OCT spectrometer, we adopted and customized the concept of a high-finesse polygon-based wavelength-scanning filter proposed in \cite{Yun:03,Oh:05,Oh:10}. Besides, various aspects of the employed approach have been also intensively considered and discussed in~\cite{10.1117/12.2282284,10.1117/12.2283034}. (quite a similar idea was introduced in~\cite{Oldenburg:03})
In order to make the system compatible and effective for mid-infrared OCT imaging, the refractive optics constituting the 4f system in the original design was replaced with reflective optics. This change was motivated by the fact that mirrors are free from chromatic aberration, which can dominate for the broad spectral range of the mid-IR OCT, disturbing performances of the spectrometer. Achromatic lenses in the mid-IR spectral region are still limited in their performances~\textendash~either in spectral bandwidth, efficiency or in aperture sizes (apertures above 1-inch are required). 
Hence, we employed standard spherical concave mirrors (gold, 100~mm focal length, tilted by 8\textdegree, SM1 and SM2 in Fig.~\ref{fig:schema}) to form the 4f system that conjugates (in the sagittal plane, where diffraction occurs) the plane of the reflective diffraction grating (300 lines/mm) and the scanner. Instead of a polygonal scanning mirror, we utilized a galvo scanner (at scanning rates from 25 up to 60~Hz, i.e. typical A-scan rates). This was done to enhance spectral resolution and reduce potential nonlinearities. In the polygonal mirror design, the rotational axis is off the facets, thus, introducing jittering (beam transitions, displacements) in the sagittal plane (i.e. spectral discrimination plane) during scanning. In our design, the rotational motion of the galvo mirror (conjugated to the grating, i.e. different spectral components hit the scanner at different angles and overlap in the plane of the scanning mirror) induces a linear spatial sweep of spectral-angular components. In other words, the scanner selects a certain wavelength over the diffraction angles by deflecting it in a specified direction.
The detailed schematic drawing of the teFD-OCT system is shown in Fig.~\ref{fig:schema}(b).

Since the spherical mirrors are used at non-normal incidence, the system exhibits astigmatism. The effects of astigmatism do not introduce any reduction of spectral resolution as they enlarge the spot size in the plane orthogonal to the scanning plane. However, they forced us to change the concept of the tunable system (initially used at the light source side to establish a swept-source using an instantaneous broadband emitter).
Hence, we utilized the spectral scanner after the OCT interferometer, i.e. as a detection system. In our configuration, each spectral component selected by the scanner is focused onto a high-sensitivity single point detector (uncooled amplified InAsSb detector; Thorlabs, PDA07P2; 1.0$\times$10\textsuperscript{-10} W/Hz\textsuperscript{1/2} noise equivalent power) behind a slit (400~\textmu m width, for scattering samples). For focusing, a long focal length spherical mirror (gold, 750~mm focal length, 200~mm away from the galvo-scanner, tilted by 19\textdegree) was used. The scanning spectrometer was designed, optimized and characterized in an optical design software (Zemax). For the given configuration of the focusing optics, beam size and the slit width, the spectral resolution of the spectrometer is 1.9~nm (with no slit inserted i.e., for the entire detector size of 0.7~mm, the resolution is 3.3~nm).
In summary, the developed spectrometer allowed us to simply implement a virtual high-resolution and high-sensitivity mid-IR linear array (equivalent physical arrays are unavailable for this spectral region) with time-separated pixels (as opposed to spatially-separated pixels of cameras).

For experimental demonstrations, primarily a high-power mid-IR supercontinuum source based on a InF\textsubscript{3} fibre (Leukos) was used. This source radiates with an average optical power of approximately 230~mW above 2.4~\textmu m wavelength (spanning approximately 4.6~\textmu m); the diameter of the outgoing beam is around 5~mm, the M\textsuperscript{2} beam quality is 1.13. The emission is pulsed (ns regime) and features a low duty-cycle. The pulse repetition rate is 250~kHz, for this reason, a boxcar gated integrator (8 cycles averaging for 25~Hz scanning rate, 4~cycles averaging for 60~Hz rate) was used to demodulate spectral interferograms and remove inter-pulse noises. An oscilloscope (Red Pitaya, STEMlab 125-14) was used to acquire signals. To illustrate the flexibility and capabilities of the developed teFD-OCT system, we also utilized a ZBLAN fiber-based supercontinuum emitter (test system, NKT Photonics), radiating only 16~mW of average optical power above 3~\textmu m wavelength at 40~kHz repetition rate (sub-ns regime, 2.7~mm beam diameter, M\textsuperscript{2}=1.09). Due to the pronounced spectral non-uniformity, a spectral bandwidth of approximately 3.4~\textmu m to 4.1~\textmu m was effectively used in this case.

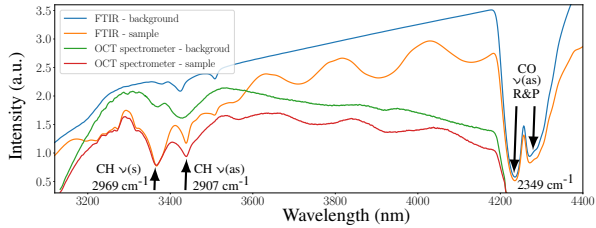
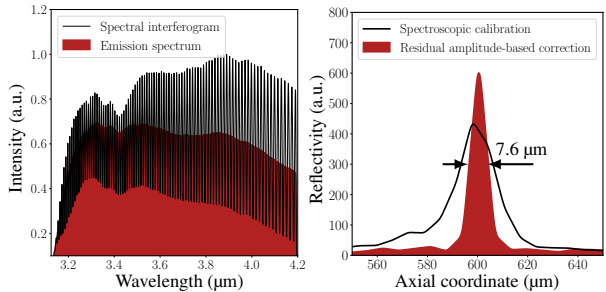
\begin{figure}[!t]
\raggedright
% \vspace{5pt}
 \begin{subfigure}[t]{1\columnwidth}
\begin{tikzpicture}
\node[anchor=south west,inner sep=0] (image) at (0.1,0,0) { \hspace{2pt} \includegraphics[width=0.9\linewidth]{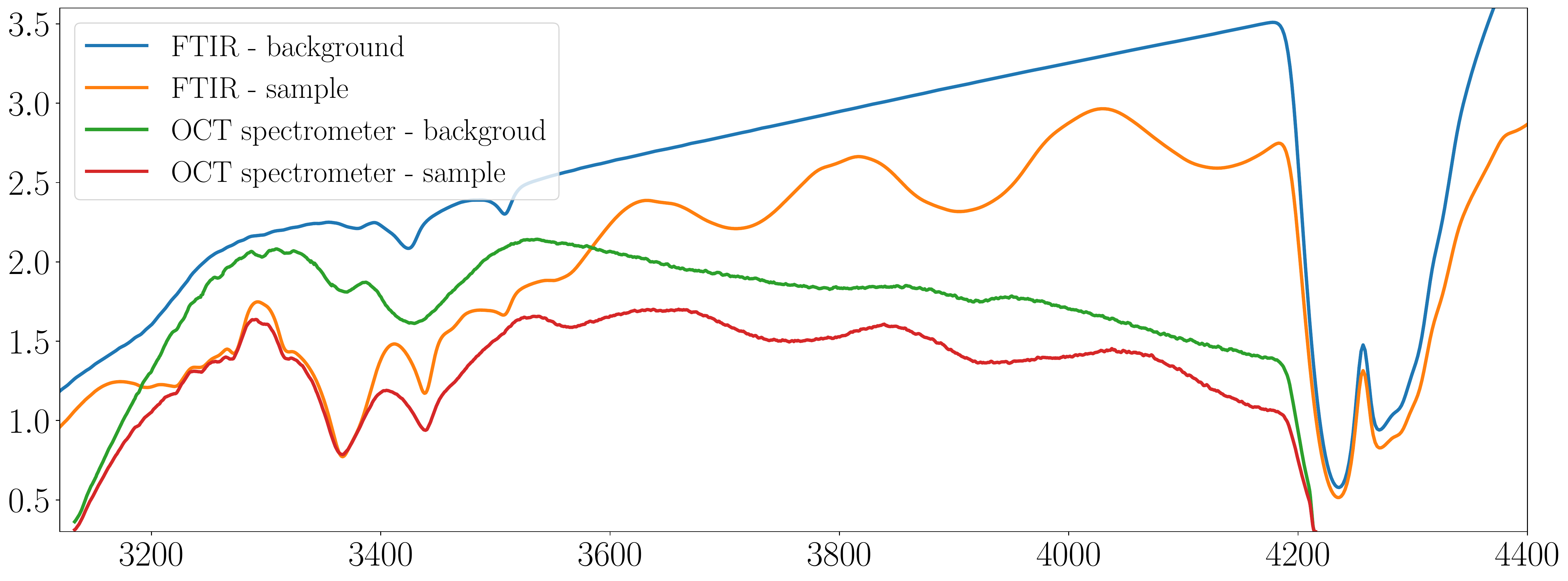}};
 \begin{scope}[x={(image.south east)},y={(image.north west)}]
  \draw (0.57,-0.025) node[]{\color{black}\scriptsize Wavelength (nm)};
  \draw (0,0.5) node[rotate=90]{\color{black}\scriptsize Intensity (a.u.)};
  \draw [-latex,thick, color={rgb,255:red,0; green,0; blue,0}] (0.24,0.1) to (.24,0.21); 
  \draw [-latex,thick, color={rgb,255:red,0; green,0; blue,0}] (0.293,0.1) to (.293,0.25); 
    \draw (0.178,0.197) node[]{\color{black}\tiny CH \textnu(s)};
    \draw (0.183,0.134) node[]{\color{black}\tiny2969 cm\textsuperscript{-1}}; 
    \draw (0.35,0.197) node[]{\color{black}\tiny CH \textnu(as)};
    \draw (0.355,0.134) node[]{\color{black}\tiny 2907 cm\textsuperscript{-1}};
     \draw [-latex,thick, color={rgb,255:red,0; green,0; blue,0}] (0.854,0.5) to (.854,0.18); 
     \draw [-latex,thick, color={rgb,255:red,0; green,0; blue,0}] (0.885,0.5) to (.885,0.29); 
     \draw (0.87,0.68) node[]{\color{black}\tiny CO};
     \draw (0.87,0.62) node[]{\color{black}\tiny \textnu(as)};
     \draw (0.87,0.56) node[]{\color{black}\tiny R\&P};
     \draw (0.91,0.14) node[]{\color{black}\tiny 2349 cm\textsuperscript{-1}};
 %% next four lines will help you to locate the point needed by forming a grid. comment these four lines in the final picture.↓
 %% upto here↑
%  %%
%  \draw[help lines,xstep=.1,ystep=.1] (0,0) grid (1,1);
%  \draw[help lines,xstep=.05,ystep=.05] (0,0) grid (1,1);
%  \foreach \x in {0,1,...,9} { \node [anchor=north] at (\x/10,0) {0.\x}; }
%  \foreach \y in {0,1,...,9} { \node [anchor=east] at (0,\y/10) {0.\y};}
 \end{scope}
\end{tikzpicture}
\caption{Spectroscopic calibration of the scanning OCT spectrometer (teSystem) using absorption bands of a PET polymer film and CO\textsubscript{2} as a reference (FTIR measurements); for the sake of clarity, raw spectra shown (spectral features used for calibration highlighted) \label{fig:calibration}}
\end{subfigure}
 \begin{subfigure}[t]{0.45\columnwidth}
\begin{tikzpicture}
\node[anchor=south west,inner sep=0] (image) at (0.1,0,0) { \hspace{2pt} \includegraphics[width=1\linewidth]{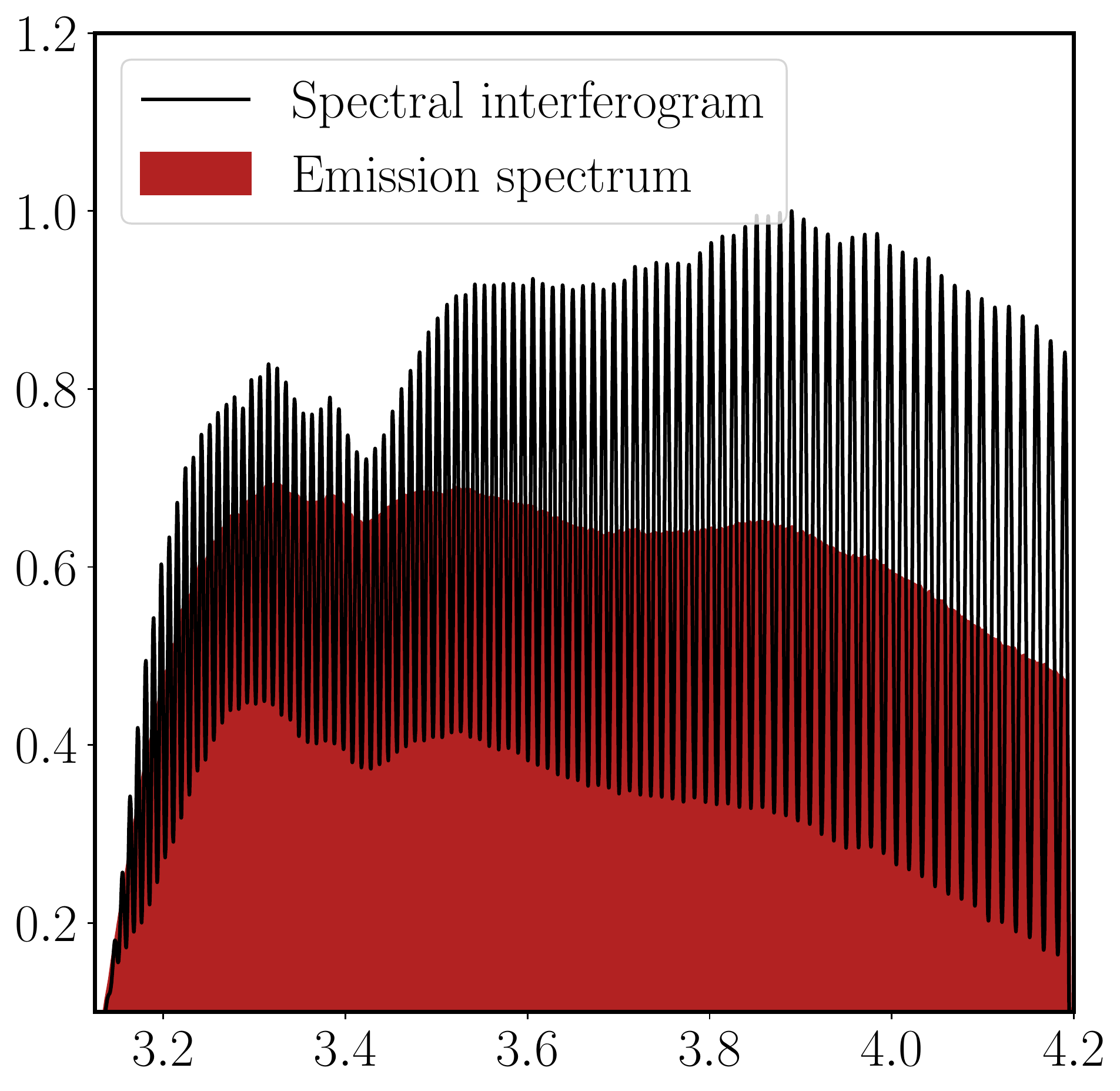}};
 \begin{scope}[x={(image.south east)},y={(image.north west)}]
  \draw (0.57,-0.025) node[]{\color{black}\scriptsize Wavelength (\textmu m)};
  \draw (0.02,0.5) node[rotate=90]{\color{black}\scriptsize Intensity (a.u.)};
 %% next four lines will help you to locate the point needed by forming a grid. comment these four lines in the final picture.↓
 %% upto here↑
%  %%
%  \draw[help lines,xstep=.1,ystep=.1] (0,0) grid (1,1);
%  \draw[help lines,xstep=.05,ystep=.05] (0,0) grid (1,1);
%  \foreach \x in {0,1,...,9} { \node [anchor=north] at (\x/10,0) {0.\x}; }
%  \foreach \y in {0,1,...,9} { \node [anchor=east] at (0,\y/10) {0.\y};}
 \end{scope}
\end{tikzpicture}
\caption{Spectrum of the InF\textsubscript{3} supercontinuum source used in OCT imaging; calibrated range (3138~nm – 4189~nm)}\label{fig:bandwidth}
\end{subfigure}
\hspace{2pt}
 \begin{subfigure}[t]{0.45\columnwidth}
\begin{tikzpicture}
\node[anchor=south west,inner sep=0] (image) at (0.1,0,0) { \hspace{2pt} \includegraphics[width=1\linewidth]{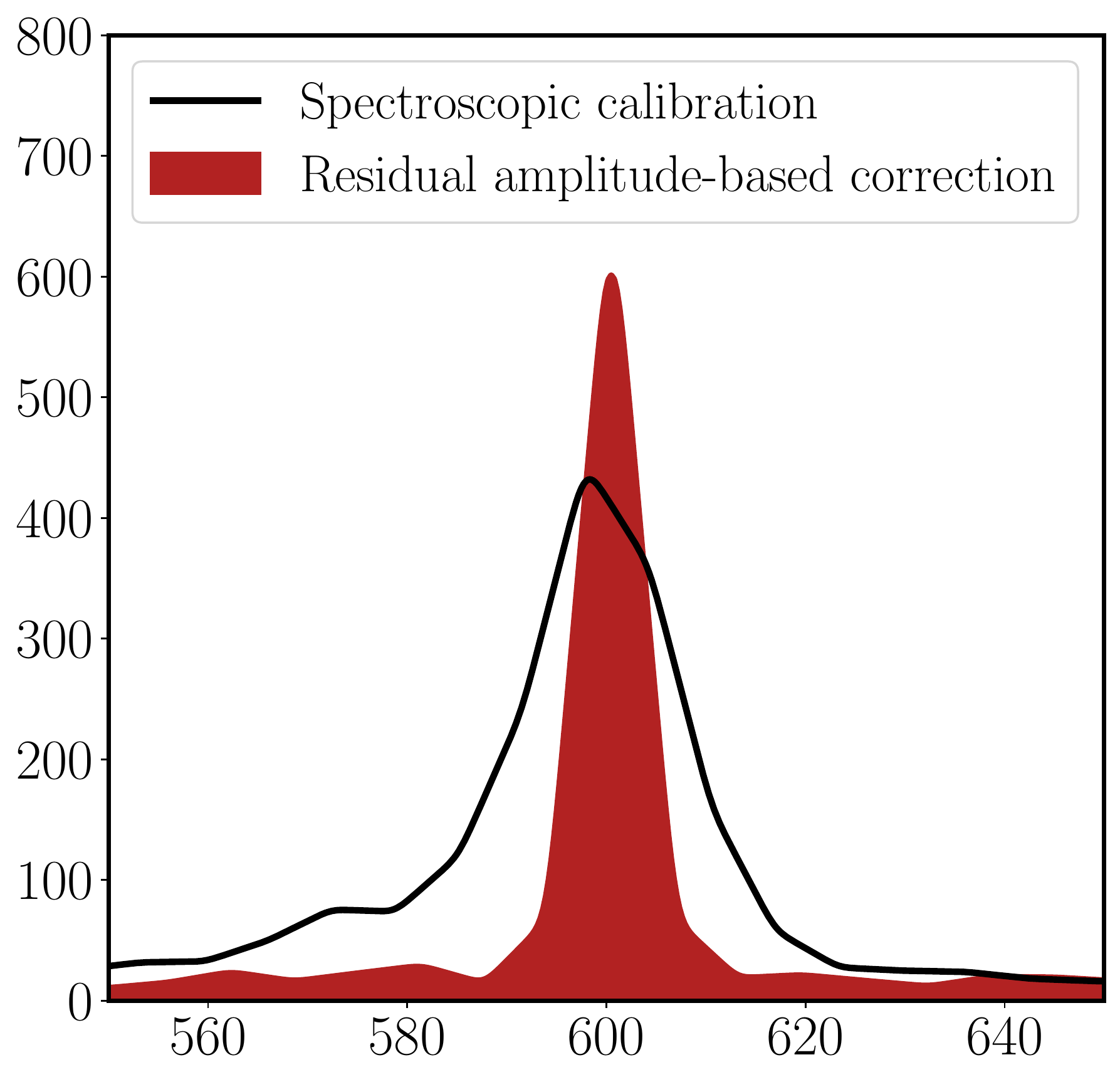}};
 \begin{scope}[x={(image.south east)},y={(image.north west)}]
  \draw (0.57,-0.025) node[]{\color{black}\scriptsize Axial coordinate (\textmu m)};
  \draw (0.03,0.5) node[rotate=90]{\color{black}\scriptsize Reflectivity (a.u.)};
\draw [-latex,thick, color={rgb,255:red,0; green,0; blue,0}] (0.72+0.025,0.41) to (.567+0.025,0.41); 
\draw [-latex,thick, color={rgb,255:red,0; green,0; blue,0}] (0.42+0.025,0.41) to (.53,0.41); 
\draw (0.705,0.46) node[]{\color{black}\scriptsize 7.6 \textmu m};
 %% next four lines will help you to locate the point needed by forming a grid. comment these four lines in the final picture.↓
 %% upto here↑
%  %%
%  \draw[help lines,xstep=.1,ystep=.1] (0,0) grid (1,1);
%  \draw[help lines,xstep=.05,ystep=.05] (0,0) grid (1,1);
%  \foreach \x in {0,1,...,9} { \node [anchor=north] at (\x/10,0) {0.\x}; }
%  \foreach \y in {0,1,...,9} { \node [anchor=east] at (0,\y/10) {0.\y};}
 \end{scope}
\end{tikzpicture}
\caption{Axial resolution characterization\label{fig:axial}}
\end{subfigure}
\caption{Performances of the teFD-OCT system: (a) calibration of the operational band; (b) accessed spectral bandwidth of mid-IR OCT (emission of the InF\textsubscript{3} fibre-based supercontinuum [Leukos]); (c) corresponding axial point spread functions.}
\label{fig:spectrum}
\end{figure}

The operating spectral band of the scanning spectrometer was calibrated prior to the remapping of spectral interferograms into \textit{k}-space (the same algorithm as in~\cite{Zorin:18} was used). For this purpose, tools provided by mid-IR spectroscopy were applied. Thus, sharp absorption peaks of polyethylene terephthalate (PET) caused by CH (carbon-hydrogen) bond stretching vibrations and absorption bands of carbon dioxide (CO\textsubscript{2}) in air were used as spectral features to get a calibration vector. Reference spectral measurements were performed with a commercial high-precision Fourier-transform infrared spectrometer (FTIR, Vertex 70, Bruker Optics). The calibration function appeared to be nonlinear, presumably due to the varying speed of the scanner at the beginning and end of a sweep cycle. However, the recorded spectra and fringes (phase and frequency) are highly time-repeatable and instabilities are not noticeable (i.e. nonlinearity does not fluctuate from sweep to sweep). Hence, the calibration procedure has been performed once and the calibration vector has been stored. The calibrated spectral window accessed by the spectrometer thus ranges from 3138~nm to 4189~nm.
Figure~\ref{fig:spectrum}(a) depicts details on the calibration procedure and spectral features used, raw spectra of the PET film (also exhibit interference patterns produced by double reflections) and background spectra (sources emissions) are shown. The measurements using the OCT spectrometer were performed with the film inserted in the reference arm, the sample arm was blocked. The calibrated emission spectrum of the InF\textsubscript{3} supercontinuum source used for OCT imaging and a typical spectral interferogram (for a single reflector) are shown in Fig.~\ref{fig:spectrum}(b).

In addition to the hardware dispersion compensation, residual dispersion effects and nonlinearities were compensated numerically using a straightforward amplitude-based approach~\cite{drexler_sleds_2015}.
Assuming prior physical knowledge of an ideal OCT signal that the fringes are equally spaced, we exploited peak positions of a reference spectral interferogram recorded for a single interface to eliminate the residual local nonlinear behaviour (similar to a zero-crossing method). The derived calibration vector was implemented into the post-processing algorithm for correction of signals in \textit{k}-space.
Figure~\ref{fig:axial} shows an obtained axial point spread function and characterization of the axial resolution (7.6~\textmu m). The sensitivity of the system based on the InF\textsubscript{3} supercontinuum source was experimentally characterized to be 80.17~dB (at 25~Hz scanning rate), using the method described in~\cite{Agrawal:17}. The teFD-OCT system based on the low-power ZBLAN source has a sensitivity of 69.3~dB and an axial resolution of 16~\textmu m.

\begin{figure}[ht]
\begin{tikzpicture}
\centering
 \begin{scope}
 %% next four lines will help you to locate the point needed by forming a grid. comment these four lines in the final picture.↓
\node[anchor=south west,inner sep=0] (image) at (0,0) {\includegraphics[width=0.53\linewidth]{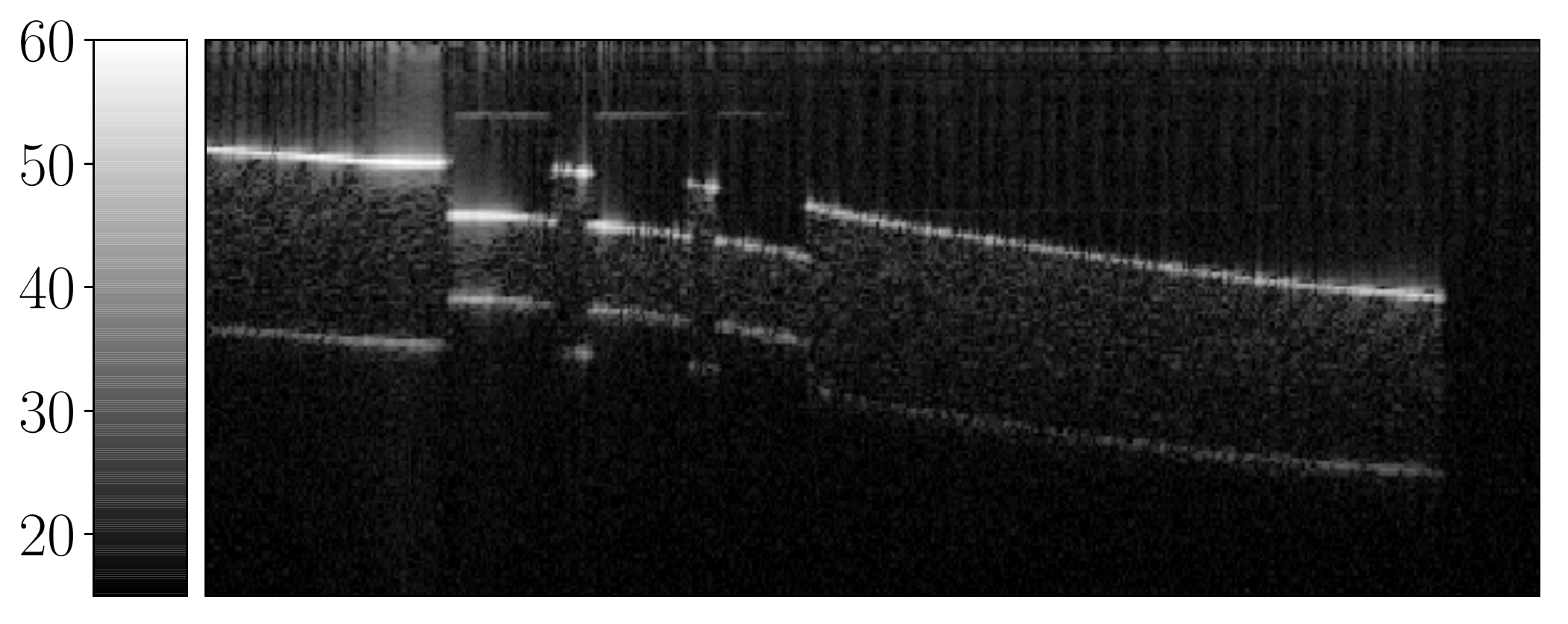}};
\node[anchor=south west,inner sep=0] (image) at (0,-1.9) {\includegraphics[width=0.536\linewidth]{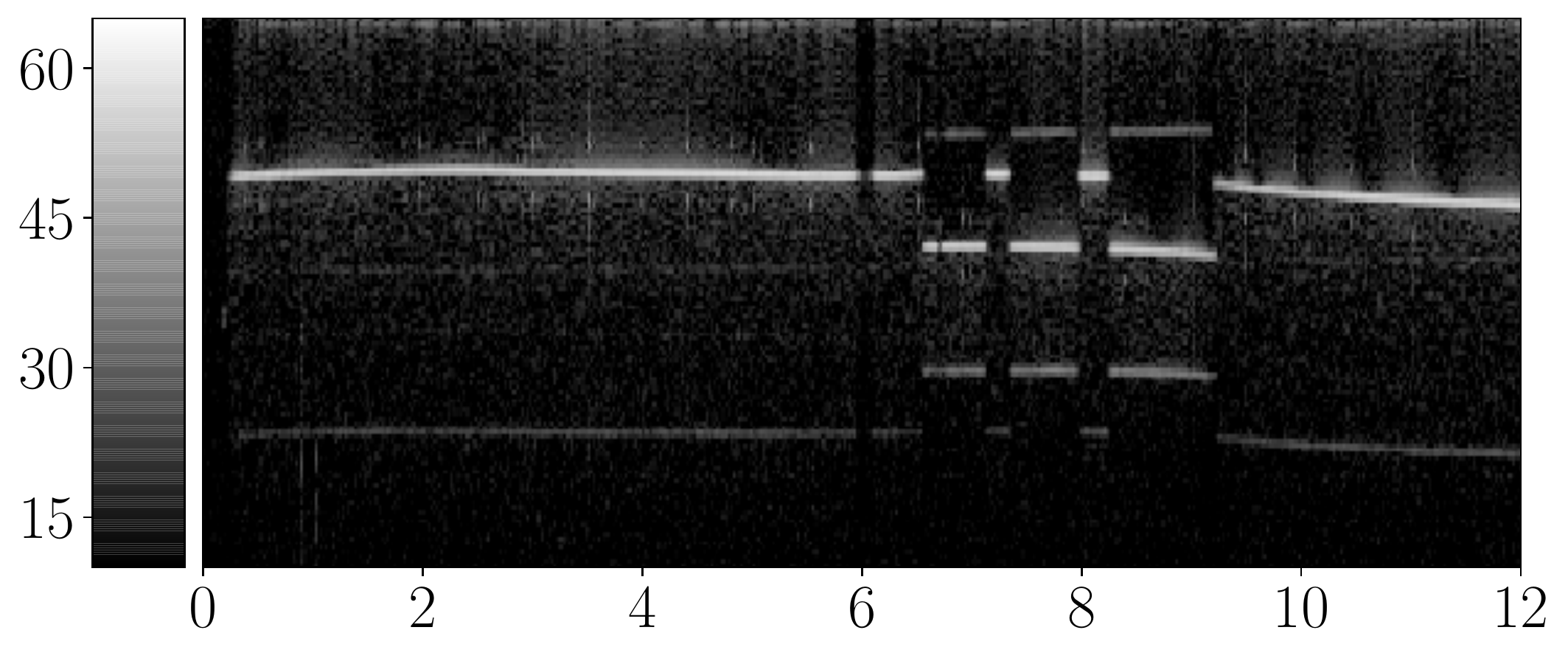}};
\node[anchor=south west,inner sep=0] (image) at (5,-1.75) {\includegraphics[width=0.45\linewidth]{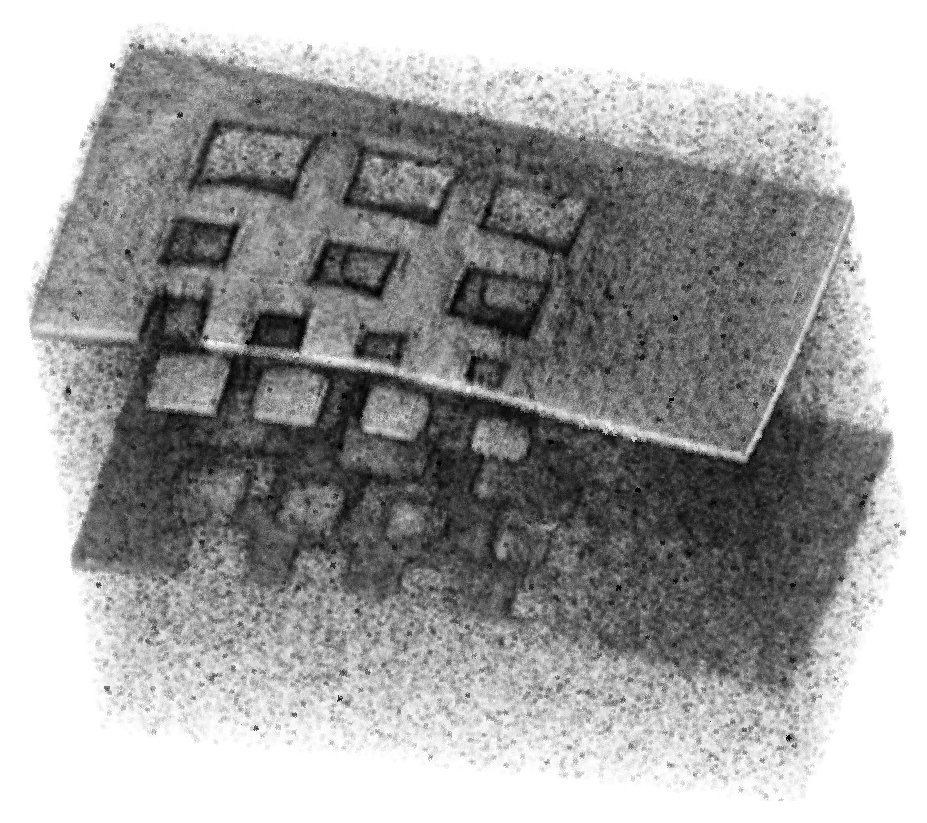}};
 \hfill
\draw [-,ultra thick, color={rgb,255:red,255; green,255; blue,255}] (0.8,0.75-0.48) to (0.8,0.75);
\draw [-,ultra thick, color={rgb,255:red,255; green,255; blue,255}] (0.8,-0.75-0.48) to (0.8,-0.75);
\draw (0.965,0.5) node[rotate=90]{\color{white}\tiny 300~\textmu m};
\draw (0.965,-1) node[rotate=90]{\color{white}\tiny 300~\textmu m};

\draw (2.4,-1.95) node[]{\color{black}\scriptsize Lateral position (mm)};
\draw (0.42,1.87) node[]{\color{black}\scriptsize dB};
\draw (4,1.45) node[]{\color{white}\scriptsize \textbf{(a)}};
\draw (4,-0.22) node[]{\color{white}\scriptsize \textbf{(c)}};
\draw (8.5,1.45) node[]{\color{black}\scriptsize \textbf{(b)}};
% %GRID
% \draw [red] (0,2) grid (9,-3);
% \draw[help lines,xstep=.5,ystep=.5] (0,2) grid (9,-3);
% \foreach \x in {0,1,...,9} { \node [anchor=north] at (\x,0) {\x}; }
% \foreach \y in {2,1,...,-2.5} { \node [anchor=east] at (0,\y) {\y}; }
 \end{scope}
\end{tikzpicture}
\caption{OCT imaging of strongly scattering (alumina, 1-2\% porosity) ceramic micro-structured test-plates: 325~\textmu m thick sample [(a) B-scan (12~sec measurement time), (b) volumetric scan (20~min measurement time); and (c) B-scan of 480~\textmu m thick plate (12~sec measurement time).}
\label{fig:testplates}
\end{figure}
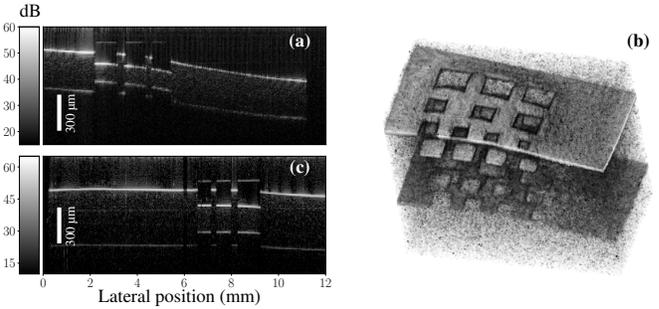

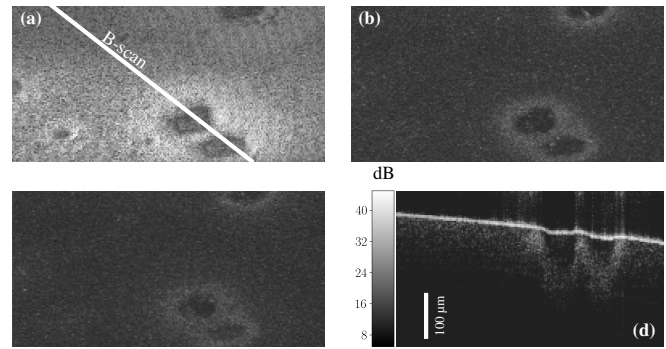
\begin{figure}[ht]
\begin{tikzpicture}
\centering
 \begin{scope}
 %% next four lines will help you to locate the point needed by forming a grid. comment these four lines in the final picture.↓
 \node[anchor=south west,inner sep=0] (image) at (0,0) {\includegraphics[width=0.4975\linewidth]{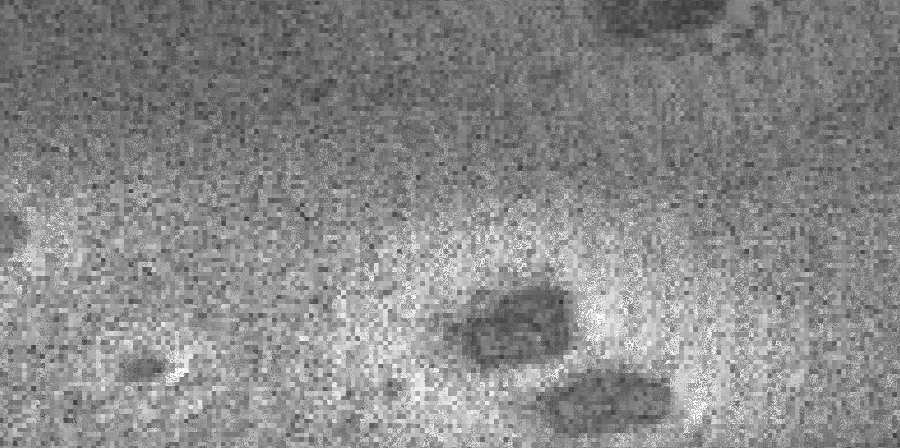}};
 \node[anchor=south west,inner sep=0] (image) at (4.5,0) {\includegraphics[width=0.4975\linewidth]{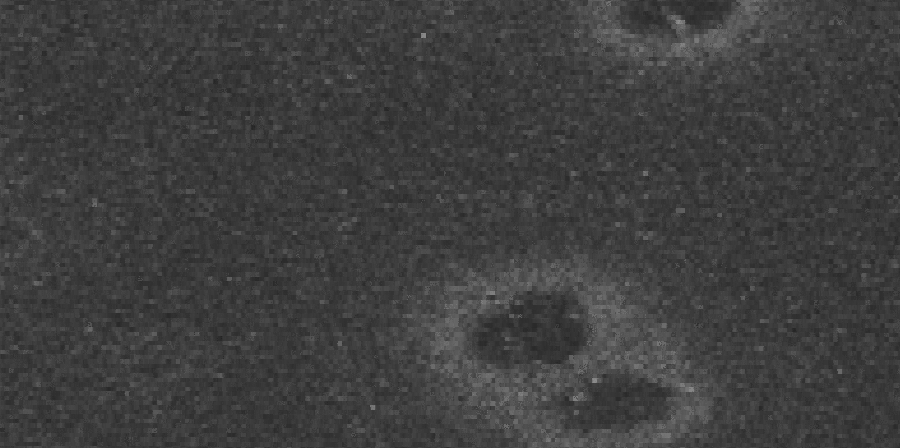}};
 \node[anchor=south west,inner sep=0] (image) at (0,-2.455) {\includegraphics[width=0.4975\linewidth]{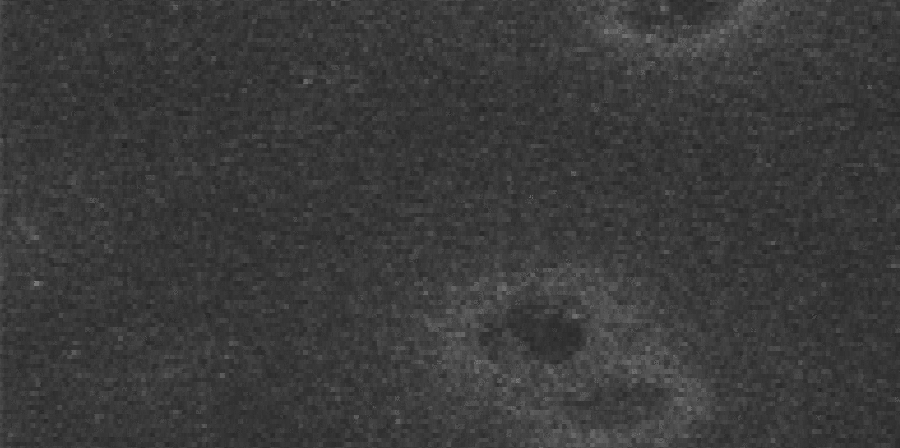}};
 \node[anchor=south west,inner sep=0] (image) at (4.5,-2.455) {\includegraphics[width=0.4975\linewidth]{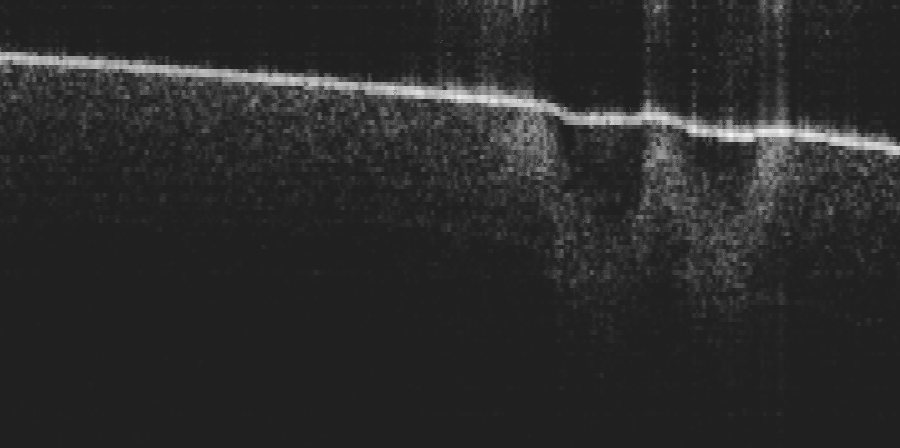}};
  \node[anchor=south west,inner sep=0] (image) at (4.5,-2.47) {\includegraphics[width=0.0698\linewidth]{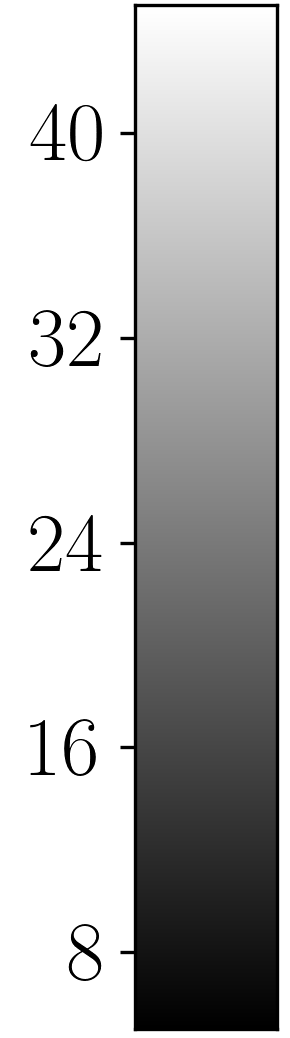}};
  \draw (4.94,-.135) node[]{\color{black}\scriptsize dB};
  \draw [-,ultra thick, color={rgb,255:red,255; green,255; blue,255}] (5.5,0.6125-2.35) to (5.5,-2.35); 
 \draw (5.7,-2.04375) node[rotate=90]{\color{white}\tiny 100~\textmu m};
 \draw (0.25,1.9) node[]{\color{white}\scriptsize \textbf{(a)}};
 \draw (4.75,1.9) node[]{\color{white}\scriptsize \textbf{(b)}};
 \draw (0.25,-0.15) node[]{\color{white}\scriptsize \textbf{(c)}};
 \draw (8.4,-2.25) node[]{\color{white}\scriptsize \textbf{(d)}};
  \draw [-,ultra thick, color={rgb,255:red,255; green,255; blue,255}] (3.2,0) to (0.5,2.1); 
    \draw (1.5,1.5) node[rotate=-35]{\color{white}\scriptsize B-scan};
 \hfill
%GRID
% \draw [red] (0,2) grid (9,-3);
% \draw[help lines,xstep=.5,ystep=.5] (0,2) grid (9,-3);
% \foreach \x in {0,1,...,9} { \node [anchor=north] at (\x,0) {\x}; }
% \foreach \y in {2,1,...,-2.5} { \node [anchor=east] at (0,\y) {\y}; }
 \end{scope}
\end{tikzpicture}
\caption{Sub-surface porosity heterogeneities for LCM-printed ceramics: (a-c) en-face images (10x5 mm\textsuperscript{2} area) at different depths (surface, about 100~\textmu m, 150~\textmu m, (25~min measurement time)); (d) B-scan, the effective depth is limited to around 0.4~mm for the alumina group index (1.83).}
\label{fig:inclusions}
\end{figure}

%\mynote{The main problem - It should be better - old measurement with lock-in and without amplitude based correction... what to do?}
%at certain technological stages of ang
The performance of the developed mid-IR OCT system was validated for structural imaging of porous industrial samples produced by an additive method of lithography-based ceramic manufacturing (LCM). Figure~\ref{fig:testplates} depicts typical OCT scans of sintered highly scattering ceramic plates (approx. 325~\textmu m and 480~\textmu m thick). The system demonstrated enhanced probing capabilities that are common for mid-IR OCT (the morphology could not be accessed by commercial near-IR OCT systems). The back interfaces of the test samples were revealed and a production curvature defect in the thin plate (325~\textmu m) was observed. 
In addition, samples showing a different class of LCM defects were examined. Figure~\ref{fig:inclusions} shows mid-IR OCT imaging of alumina ceramics (same material properties) with local subsurface porosity variations. The presence of short mid-IR wavelengths in the broad OCT bandwidth allowed us to gain some advantage and visualize varying scattering behaviour. Thus, depth profiles of clusters with reduced porosity were accessed. High porosity zones surrounding the clusters could be identified.

\begin{figure}[ht]
\centering
\begin{tikzpicture}
 \begin{scope}
 %% next four lines will help you to locate the point needed by forming a grid. comment these four lines in the final picture.↓
\node[anchor=south west,inner sep=0] (image) at (0,0) {\includegraphics[width=0.9\linewidth]{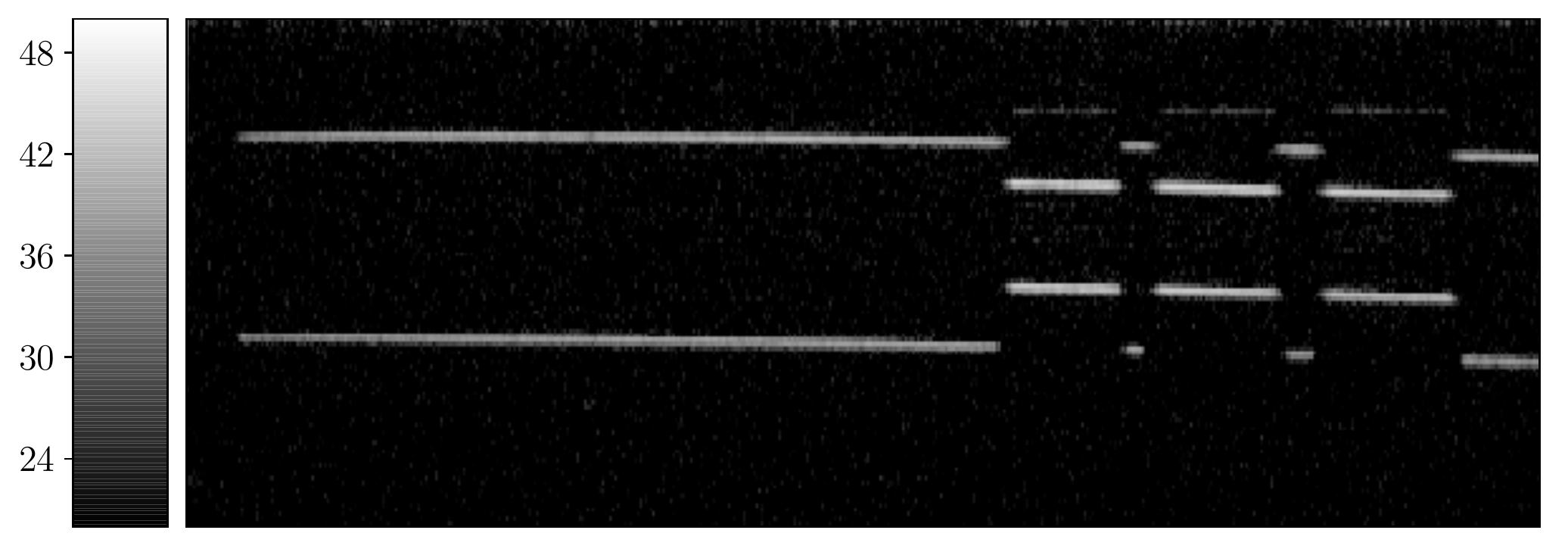}};
\node[anchor=south west,inner sep=0] (image) at (0,-2.9) {\includegraphics[width=0.905\linewidth]{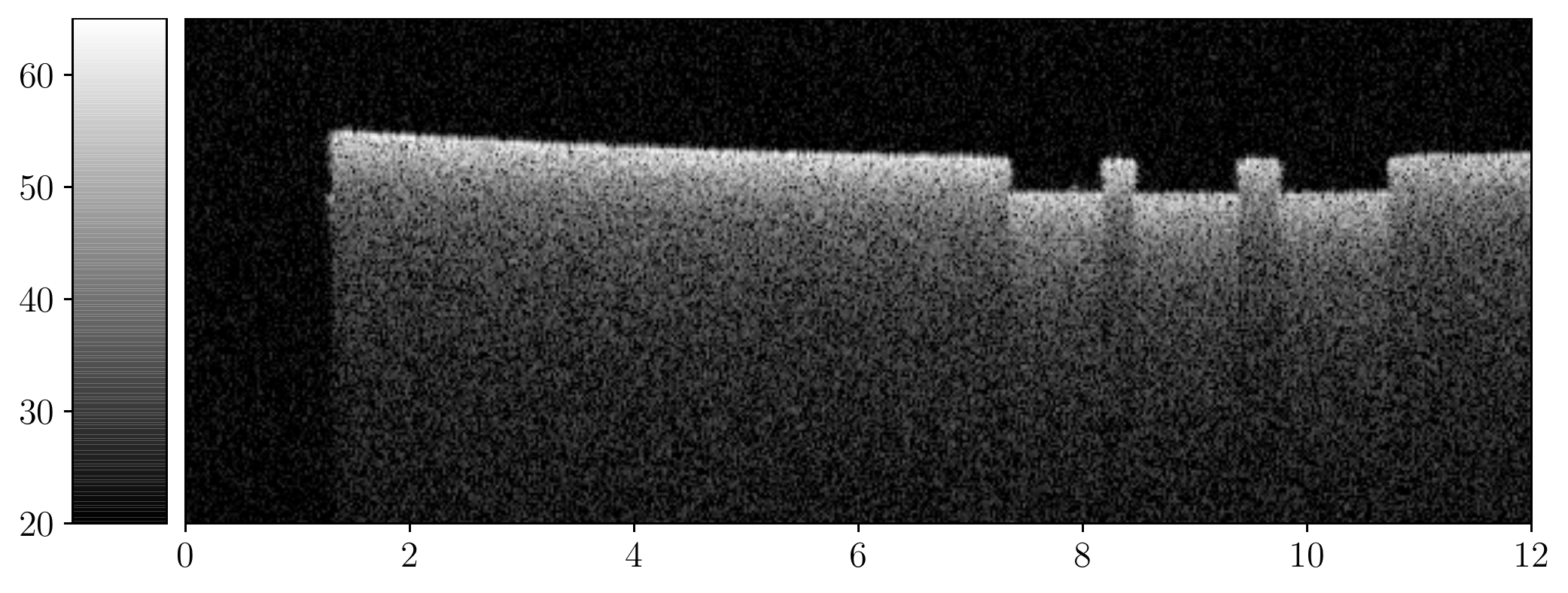}};
% \node[anchor=south west,inner sep=0] (image) at (5,-1.75) {\includegraphics[width=0.45\linewidth]{cscan_plates.png}};
%  \hfill
\draw [-,ultra thick, color={rgb,255:red,255; green,255; blue,255}] (1.2,0.9) to (1.2,0.9-0.508);
\draw [-,ultra thick, color={rgb,255:red,255; green,255; blue,255}] (1.2,-1.75) to (1.2,-1.75-0.508);
\draw (1.35,0.65) node[rotate=90]{\color{white}\tiny 300~\textmu m};
\draw (1.35,-2) node[rotate=90]{\color{white}\tiny 300~\textmu m};
% \draw (2.4,-1.95) node[]{\color{black}\scriptsize Lateral position (mm)};
\draw (0.6,2.85) node[]{\color{black}\scriptsize dB};
\draw (4.2,-2.9) node[]{\color{black}\scriptsize Lateral position (mm)};
\draw (1.35,2.3) node[]{\color{white}\small \textbf{(a)}};
\draw (1.35,-0.4) node[]{\color{white}\small \textbf{(b)}};
% \draw (8.5,1.55) node[]{\color{black}\scriptsize (c)};
% %GRID
% \draw [red] (0,2) grid (9,-3);
% \draw[help lines,xstep=.5,ystep=.5] (0,2) grid (9,-3);
% \foreach \x in {0,1,...,9} { \node [anchor=north] at (\x,0) {\x}; }
% \foreach \y in {2,1,...,-2.5} { \node [anchor=east] at (0,\y) {\y}; }
 \end{scope}
\end{tikzpicture}
\caption{OCT imaging of the green LCM parts using (a) the mid-IR teFD-OCT with the 16~mW supercontinuum source (15~sec measurement time) and (b) the commercial 1.3~\textmu m SS-OCT system.}
\label{fig:nkt}
\end{figure}
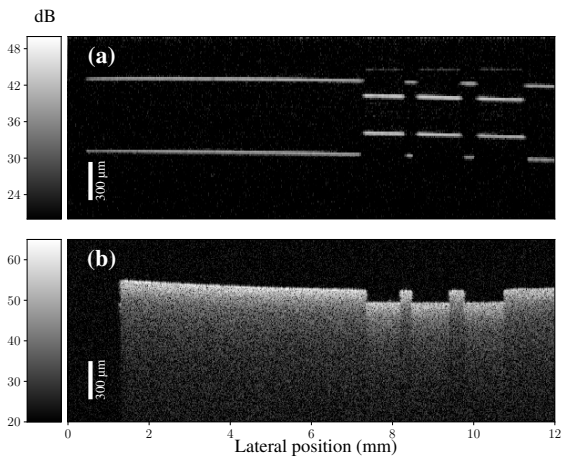

Particularly interesting results were obtained for imaging of green LCM parts (intermediate production stage) using a 16~mW mid-IR supercontinuum source. The samples are strongly scattering, and made of a solid, flexible material that consists of ceramic particles embedded in a polymer matrix (on the basis of acrylate and methacrylate). A B-scan of the 650~\textmu m thick green sample [corresponds to the sintered part shown in Fig.~\ref{fig:testplates}(b)] is depicted in Fig.~\ref{fig:nkt}(a). For comparison, a B-scan obtained using a state-of-the art commercial near-IR OCT (1.3~\textmu m center wavelength, swept-source OCT, Vega, Thorlabs) is shown in Fig.~\ref{fig:nkt}(b). The recorded scans demonstrate the enhanced probing performance but also perspectives for the mid-IR OCT based on the low-power light source; for the near-IR OCT system multiple scattering dominates, the bottom interface is not detected.

In conclusion, we have proposed and experimentally demonstrated a concept of mid-IR time-encoded FD-OCT operating in direct sensing mode. The developed system combines the strengths of the TD- and FD-OCT configurations as it employs a single pixel detector, but also inherits the specific sensitivity advantage by exploiting the concept of a virtual array with time-separated pixels. The experimental results have indicated practical prospects for applications in field conditions. As future improvements, we foresee the optimization of imaging speed and sensitivity. We expect further adaptation of the optical scheme to enhance spectral performances and see a possibility to establish kinetic mid-IR sweep OCT sources based on a similar design.% for mid-IR OCT applications.

\vspace{5pt}
\textbf{Funding }{\"O}sterreichische Forschungsf{\"o}rderungsgesell-schaft (FFG) (877481, 856896); State of Upper Austria (Wi-2020-700476/3).\newline
\textbf{{Acknowledgments }}The authors thank Guillaume Huss from Leukos and Patrick Bowen from NKT Photonics for providing supercontinuum sources; and Martin Schwentenwein and Dominik Brouczek from Lithoz for designing and providing us with various LCM ceramic samples.\newline
% \bmsection{Disclosures}The authors declare no conflicts of interest.
% \bmsection{Data availability}Data underlying the results presented in this paper are not publicly available at this time but may be obtained from the authors upon reasonable request.
% \end{backmatter}

\bibliographystyle{ieeetr}
\bibliography{refs}
\end{document}